# Magnetic Transition in the Kondo Lattice System CeRhSn$_2$

Z. Hossain[1], L.C. Gupta[2] and C. Geibel[1]

[1]Max-Planck Institute for Chemical Physics of Solids, Nöthnitzer Str. 40, 01187 Dresden, Germany.
[2]Tata Institute of Fundamental Research, Homi Bhabha Road, Mumbai 400 005, India

**Abstract**
We present the results of our measurements of magnetization, heat capacity, resistivity and magnetoresistance of the Ce-based intermetallic compound CeRhSn$_2$. The magnetic transition, as reported earlier [1], is confirmed by the presence of well defined anomalies in magnetization, resistivity and specific heat. The Ce-ions are in a crystal field doublet ground state and undergo a magnetic transition below ~ 4 K. The magnetic contribution to the resistivity is similar to that of a Kondo-lattice system which orders magnetically. The magnetoresistance of the compound measured at 2K in a field of 10 T is negative and large in magnitude (- 41%).

**Introduction**

Among the rare-earth compounds, Ce-based intermetallics have been of special interest as several of them exhibit anomalous physical properties. Competition between the Kondo and the RKKY interactions leads to a variety of exotic ground states such as magnetic Kondo lattice with reduced magnetic moment, heavy fermion state (both normal and superconducting) and Kondo insulator etc [2]. Over the past few years, there has been great interest in antiferromagnetic Ce-compounds, such as CePd$_2$Si$_2$, CeRh$_2$Si$_2$ and CeIn$_3$ because their $T_N$'s are highly sensitive to applied pressure and eventually at certain pressure, via a heavy fermion state, they undergo a superconducting transition at low temperatures (~ a few hundred mK) [3-6]. This superconducting state is believed to be characterized by *magnetically mediated Cooper pairing*. Very recently, superconductivity has also been observed in UGe$_2$ below 1 K, in a limited pressure range on the border of ferromagnetism [7]. Thus it is of immediate importance to investigate the response to the applied pressure of Ce-based compounds that undergo ferromagnetic ordering. It is in this regard that the title compound CeRhSn$_2$ acquires significance.

The new material CeRhSn$_2$ has recently been reported to form in an orthorhombic structure (space group: *Cmcm*) and its magnetic susceptibility, magnetization (down to 2 K), resistivity (down to 4K) and $^{119}$Sn Mössbauer have been studied [1]. From the temperature dependence of its magnetic susceptibility, it was inferred that CeRhSn$_2$ undergoes a ferro(ferri)magnetic transition at ~ 4K. Further, a resistivity minimum observed around 30 K in CeRhSn$_2$ was conjectured to be arising due to a Kondo type interaction. Thus, this material seems to have the right kind of ingredients required for a truly exciting ground state at high pressure and low temperatures. Considering that *there are not many* ferro(ferri) magnetic Kondo compounds, we have planned to investigate CeRhSn$_2$ extensivley. In the present work, besides reconfirming the magnetic ordering, we have further characterized the ground state of CeRhSn$_2$.

Polycrystalline samples of CeRhSn$_2$ were prepared by arc melting stoichiometric amounts of high purity elements on a copper hearth under argon atmosphere. Weight loss during the melting process was negligible. The samples were annealed at 900 ºC for one week. Powder x-ray diffraction showed that the samples were single phase, crystallizing in the reported structure (orthorhombic structure, space group: *Cmcm*) with lattice parameters a =4.5929 (6) Å, b = 16.9849 (24) Å and c = 9.5896 (14) Å which are in agreement with the published crystallographic data [1]. Resistivity and magnetoresistance were measured in the temperature range 2 K – 300 K and in magnetic fields up to 10 T using a standard low frequency ac four-probe technique. Magnetization measurements were performed, 2K < T < 300 K, using a commercial SQUID magnetometer (Quantum Design). Heat capacity measurements, 0.35 K < T < 25 K, were carried out in a commercial Physical Property Measurement System (Quantum Design) using the relaxation method.

Magnetic susceptibility $\chi(T)$ exhibits a Curie-Weiss behavior in the temperature range 50 K – 300 K with an effective magnetic moment $\mu_{eff}$ ~ 2.4 $\mu_B$, close to the moment expected for a free trivalent Ce ion, and a paramagnetic Curie temperature $\theta_p$ ~ –11 K. Figure 1 shows the magnetization M(T) below 15 K, measured in several applied fields. It is clearly seen that the temperature dependence of M(T) is strongly field dependent below ~ 5 K. The top panel of Fig. 1 shows M(T) for B = 2 mT both in the field cooled as well as zero field cooled configurations. The material undergoes two closely spaced transitions at $T_{M1}$ = 4 K (which was reported in Ref.1) and $T_{M2}$ = 3 K. At B = 20 mT, 0.1 T and 3 T (see the three bottom panels), one does not see the higher temperature transition ($T_{M1}$ = 4 K) and M(T) shows a tendency of saturation below 3 K. Low temperature neutron diffraction measurements should throw light on the subtleties of the multiple magnetic transitions.

The magnetization M(B) of the sample, measured at 2 K (*below $T_{M2}$*) in applied magnetic fields up to 5 T, is shown in Fig. 2. For comparison, M(B) measured at 10 K (*above $T_{M1}$*) is also shown in Fig. 2. In the magnetically ordered region (T = 2 K), an upturn in the M(B) vs. B curve is seen at B ~ 0.2 T. The magnetic moment at 2 K and 5 T is 0.9 $\mu_B$/Ce, much less than the saturation moment for the free trivalent Ce ions. This value is higer than ~ 0.75 $\mu_B$/Ce reported in Ref. 1 and the difference is due to preferred orientation of the polycrystalline samples. We find significant difference in the value of the magnetization when measured along two perpendicular directions. Measurements on single crystalline samples are required to determine the anisotropic magnetic property of this compound.

The specific heat ($C_p$) data show a strong anomaly at ~ 3.5 K due to magnetic ordering (Fig. 3). The value of $C_p$ at the peak is ~ 7 J/mol K and the entropy corresponding to the magnetic transition is ~ 0.85$R$ln2, which would suggest that the ground state crystal field doublet is well separated from the first excited state. The resistivity data also suggest a crystal field splitting of about 100 K. Below 1 K (T << $T_{M1},T_{M2}$) the specific heat follows $C_p$ = $\gamma T+\beta T^3$ expected for an antiferromagnet rather than $\gamma T+\beta T^{3/2}$ expected for a ferromagnet. However, since the low field magnetization shows a rapid increase below 5 K, the magnetic structure is not a simple antiferromagnet, rather it has to be one with a considerable amount of ferromagnetic component such as canted antiferromagnetic or ferrimagnetic type. A possibility for the latter scenario is given by the presence of two inequivalent Ce sites in the crystal structure, which could have slightly different crystal-field ground state magnetic moments. The $\gamma$ value obtained from this fit is ~ 300 mJ/mole K$^2$. It is not clear at present whether this comparatively large $\gamma$ value is due to a pure electronic contribution, thus indicating a large effective mass, or whether it is partly due to low-lying magnetic excitations.

The room temperature value of the electrical resistivity $\rho$(T = 300 K) ~ 100 $\mu\Omega$ cm is typical of Ce-based intermetallic compounds and the value is close to the value reported in Ref 1. The temperature dependence of $\rho$(T) for T > 25 K suggests metallic character ($\rho$(T) decreases with decrease of temperature) of the material. At low temperatures ( ~ 5 K < T < 20

K), however, the resistivity increases with decreasing temperature (Fig. 4). The magnetic part of the resistivity (inset Fig. 4), which is obtained by subtracting the resistivity of LaRhSn$_2$ taken from Ref. 1, behaves as –lnT in two temperature regions separated by a maximum at ~ 100 K. This kind of behavior of the magnetic resistivity is typical of Ce-based Kondo-lattice compounds in the presence of a crystal-field [8]. The temperature at which the maximum occurs is related to the crystal-field splitting which is about 100 K in the present case. The shape of the magnetic resistivity is similar to that expected for a crystal-field doublet ground state [see Fig. 1 in Ref. 8]. Below 4 K the resistivity shows a sharp drop due to the magnetic transition.

We have also measured the resistivity as a function of applied magnetic field at two fixed temperatures 2 K and 10 K; the results are shown in Fig. 5. The magnetoresistance ($\rho(B,T) - \rho(0,T))/ \rho(0,T)$ at T = 10 K and B = 10 T is negative and rather large in magnitude (–15.8%). In the paramagnetic region the negative magnetoresistance is due to the freezing out of spin flip scattering in a Kondo compound by the magnetic field. In the magnetically ordered phase, at 2 K, $\rho(B)$ shows initially a rapid *decrease* as B *increases*. On further increase of B, a broad maximum, centred at ~ 3.5 T, is observed. The magnetoresistance at 2 K (in the magnetically ordered phase) and 10 T, is negative and quite high in magnitude (– 41 %). While the negative magnetoresistance in the magnetically ordered region is due to field-induced ferromagnetism the broad maximum around 3.5 T suggests a complicated magnetic phase diagram.

To summarize, our resistivity, magnetoresistance, magnetization and specific heat data provide unambiguous evidence that CeRhSn$_2$ is a Kondo lattice compound which undergoes magnetic transition(s) below 4 K. Magnetization measurements reveal the presence of a metamagnetic transition. At present we are not able to clearly state the nature of the magnetic order, except the fact that it is not a ferromagnetic or collinear antiferromagnetic. Ferrimagnetic or canted antiferromagnetic might be a possible type of magnetic order. Neutron scattering studies should be carried out to obtain a good understanding of the magnetism of CeRhSn$_2$. Also, it would be rewarding to determine the influence of pressure on the magnetic behavior of this compound.

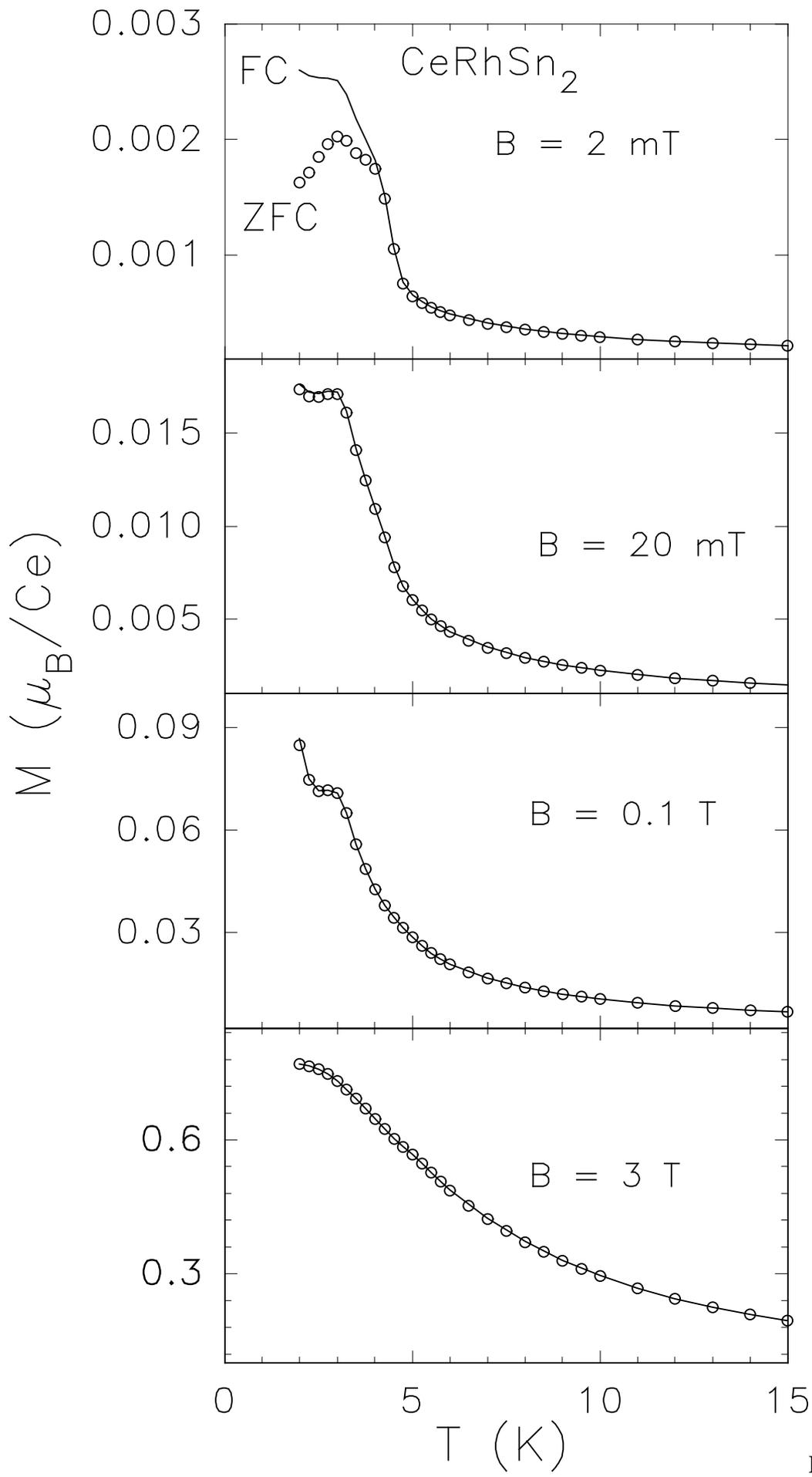

Fig. 1

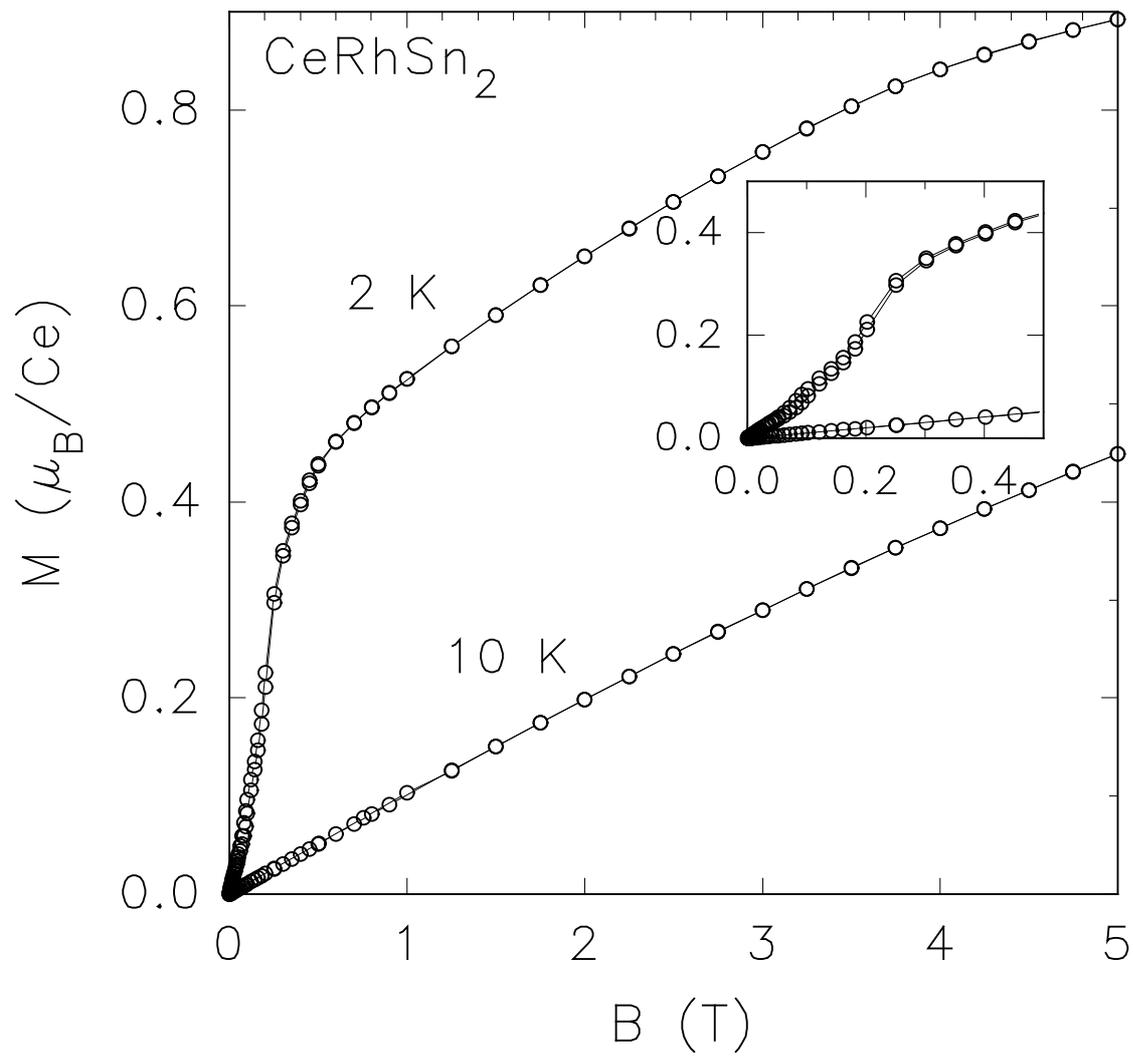

Fig.2. Magnetization as a function of field at 2 K and 10 K.

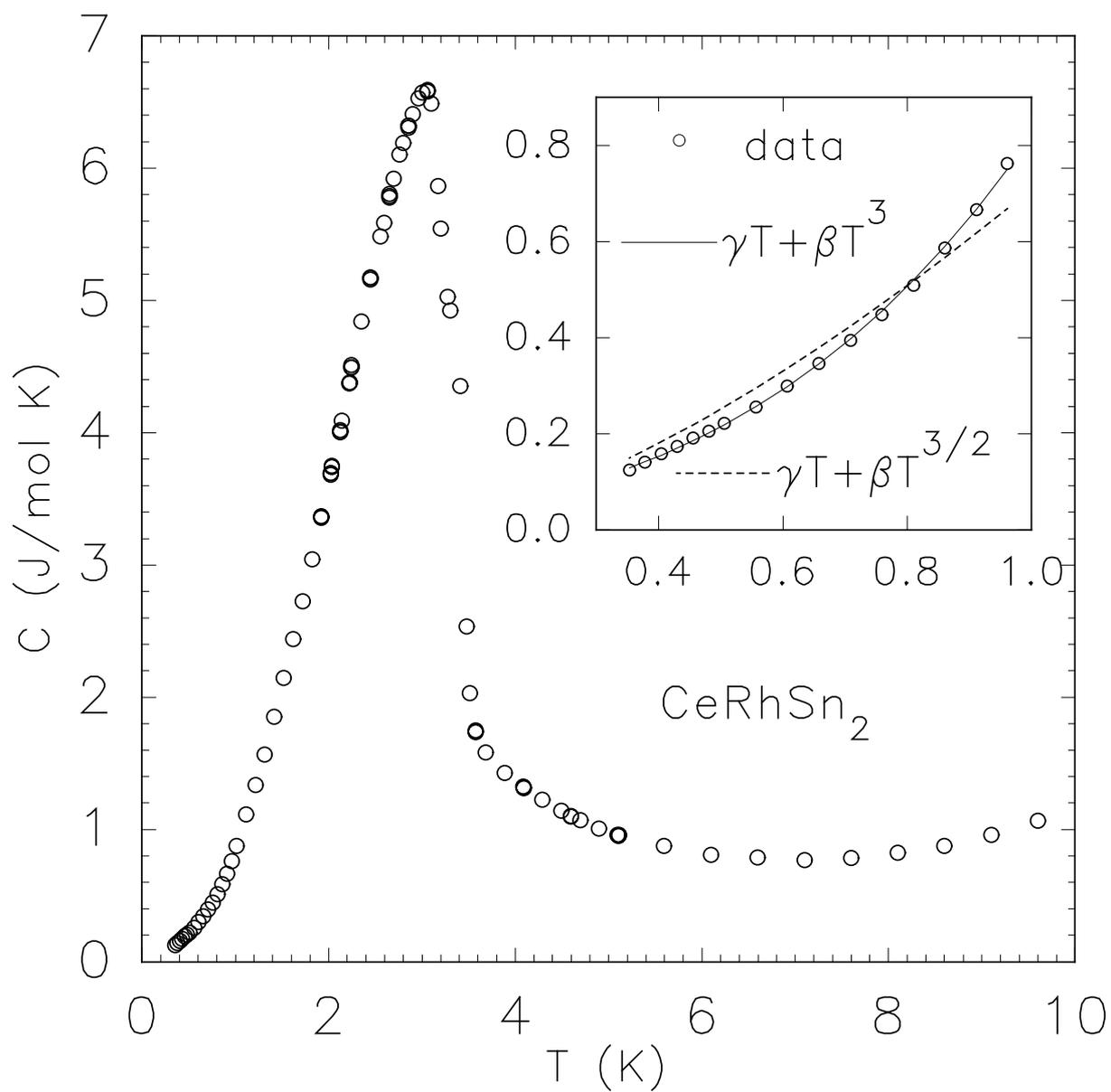

Fig.3. Specific heat of CeRhSn$_2$. The inset shows fits of the experimental data as mentioned in the text.

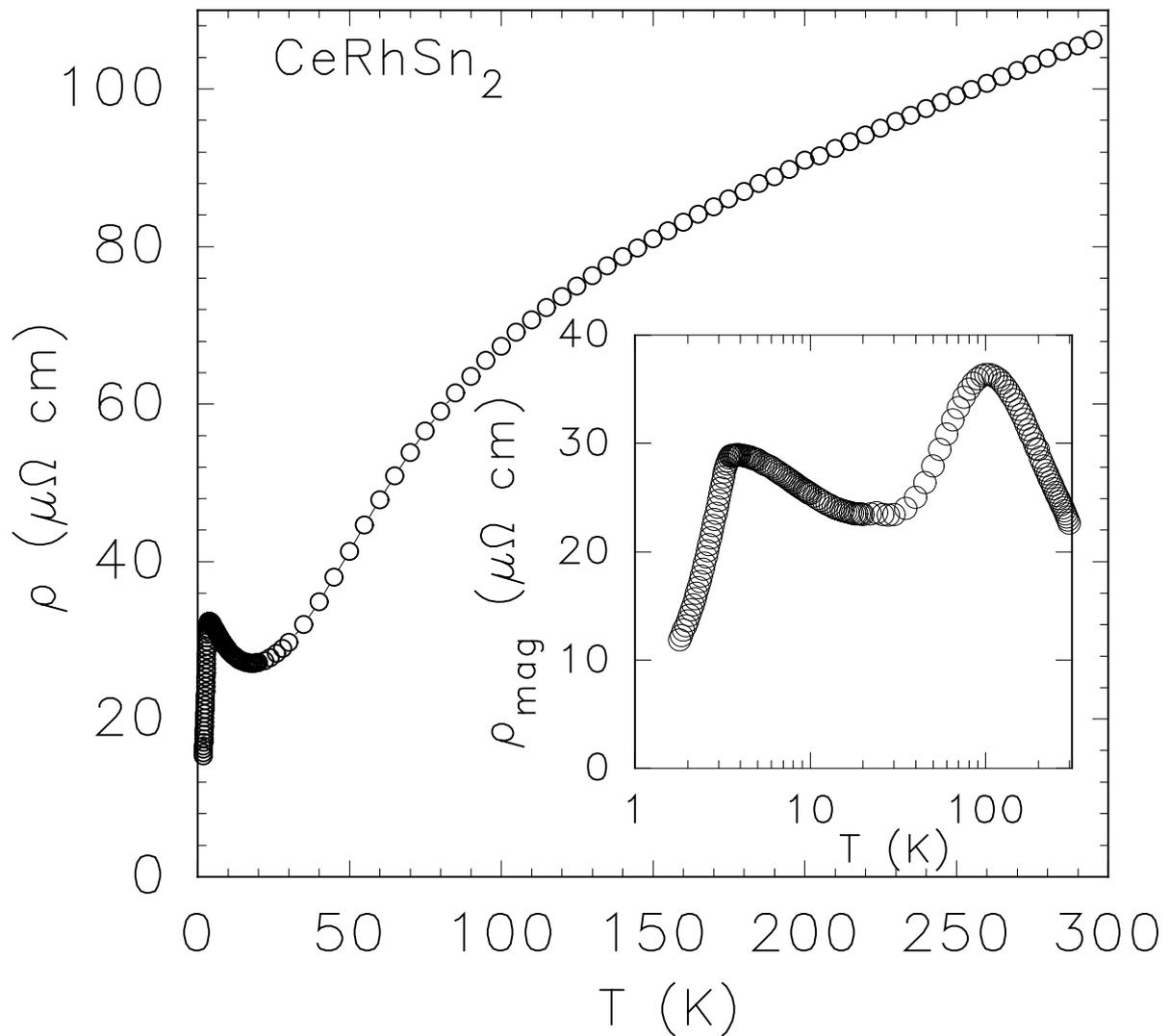

Fig.4. Resistivity (ρ) of CeRhSn$_2$ as a function of temperature. The inset shows the magnetic part of the resisvity (ρ$_{mag}$).

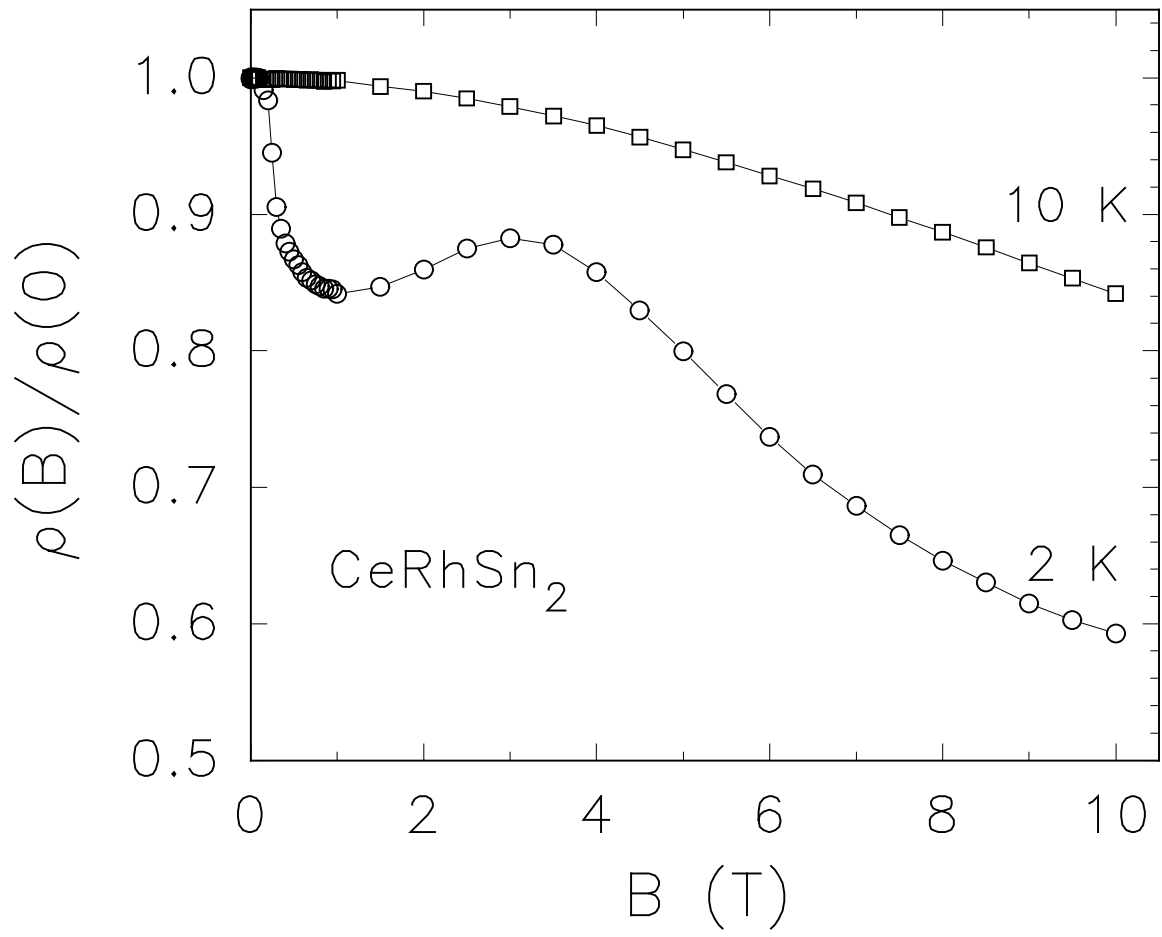

Fig.5. Normalized magnetoresistance as a function of field for CeRhSn$_2$.